\documentclass[aps,twocolumn,showpacs,showkeys]{revtex4}   
\usepackage{epsfig}   
   
\begin{document} 
   
\title{Obtaining pressure versus concentration phase diagrams in spin 
  systems from Monte Carlo simulations}  
  
\author{Carlos E. Fiore} 
 
\email{fiore@if.usp.br} 
 
\author{C. E. I. Carneiro} 
 
\email{ceugenio@if.usp.br} 
   
\affiliation{Instituto de F\'{\i}sica\\   
Universidade de S\~{a}o Paulo\\   
Caixa Postal 66318\\   
05315-970 S\~{a}o Paulo, S\~{a}o Paulo, Brazil}   
\date{\today}   
   
\begin{abstract}  
  
We propose an efficient procedure 
for determining phase diagrams of systems that are described by spin  
models. It consists of combining cluster algorithms with the method 
proposed by Sauerwein and de Oliveira where the grand canonical potential    
is obtained directly from the Monte Carlo simulation, without the necessity of 
performing numerical integrations. The cluster algorithm presented  
in this paper eliminates metastability in first order phase transitions  
allowing us to locate precisely the first-order transitions lines. We also 
produce a different technique for calculating the thermodynamic limit
of quantities such as the magnetization whose infinite volume limit is not 
straightforward in first order phase transitions. As an application, we 
study the Andelman model for Langmuir monolayers made of chiral 
molecules that is   
equivalent to the Blume-Emery-Griffiths spin-1 model. We have obtained the  
phase diagrams in the case where the intermolecular forces favor  
interactions between enantiomers of the same type (homochiral  
interactions). In particular, we have determined diagrams in the     
surface pressure versus concentration plane which are more relevant from  
the experimental point of view and less usual in numerical studies.    
\end{abstract}  
 
 
  
\maketitle   
  
  
\section{Introduction}   
   
The importance of numerical simulations in Physics is due to the fact  
that very few models can be exactly solved. In principle one may directly  
simulate any model on a computer.  
Moreover, the Metropolis \cite{metr} and the Glauber \cite{glauber} 
algorithms used in Monte Carlo (MC) simulations are very general and 
easy to implement. In practice things are not so simple. Near 
second-order phase transitions the configurations generated by these 
algorithms present strong temporal   
correlations (critical slowing down), which prevent an efficient sampling  
of the configuration space. In addition, hysteresis effects due to  
metastability prevent a precise location of first-order transition lines. 
  
In the last years, several techniques have been proposed to circumvent  
these problems such as the reweighting technique by Berg and Neuhaus
\cite{berg} and  
the simulated tempering by Marinari and Parisi \cite{parisi}. A
different approach is the   
use of cluster algorithms pioneered by Swendsen and Wang \cite{sw} and by  
Wolff \cite{wolff}.         
Several studies  have shown the efficiency of cluster algorithms in  
reducing the critical slowing down \cite{sw}. More recently, it has been
shown that the cluster dynamics   
may practically eliminate metastability in first order phase transitions
So far this has been achieved only for the Blume-Emery-Griffiths (BEG)
spin-1 model \cite{BEGMODEL,BEG2}, for which a special cluster algorithm
has been developed \cite{bouabci,rachadi}.   
  
In this paper, we present a simple cluster algorithm that  
eliminates metastability in first-order phase transitions and combine  
it to the Sauerwein and de Oliveira (SO) method that allows us to obtain the 
surface pressure directly from numerical simulations \cite{sauerwein}. In 
the original formulation of the SO method, the authors used the Metropolis   
dynamics to generate the system configurations. However, as explained above,  
this is not the best choice near phase boundaries. We also introduce a
simple procedure to calculate the order parameters and the concentrations
of molecules in the neighborhood of a first-order line from numerical
simulations.   

As an application of our method, we have determined the phase  
diagrams of the Andelman  model for Langmuir monolayers made of  
chiral molecules. More specifically, we were interested in surface pressure 
versus concentration phase diagrams, which are more interesting from the  
experimental point of view. The Andelman model was first studied by using
the mean  field approach on a bipartite lattice \cite{andelman1}. However,
X-ray diffraction experiments suggest that the condensed phases of Langmuir 
monolayers tend to form triangular structures, and not a bipartite lattice as 
considered by Andelman. In order to be more consistent with the physics   
of Langmuir monolayers, Pelizzola et al \cite{pelizzola}    
have studied the heterochiral case on a two dimensional triangular lattice,   
using the cluster variation method, and have obtained  
phase diagrams that are qualitatively different from Andelman's.   
We study through MC simulations the remaining homochiral case, and we show 
that, in contrast with the heterochiral case, the MC and the mean field 
methods give results that are in good agreement.     
  
This paper is organized as follows: in section 2 we  
briefly review the Andelman  model, in section 3 we present the  
cluster algorithm and briefly review the Sauerwein de Oliveira method, in 
section 4 we discuss the numerical results and conclude in section 5.  
  
  
\section{Chiral Langmuir monolayers and the 
Blume-Emery-Griffiths  model}    
  
Langmuir monolayers are formed by spreading amphiphilic    
molecules in an  air-water interface. Amphiphilic molecules are strongly    
asymmetric, constituted by two parts with opposite features. The first  
part---the head---,  is  hydrophilic. It is  made of polar chemical groups
and remains on the water. The second part---the tail---, is hydrophobic and  
made of hydrocarbon chains which remain in the air.   
When the tail is strongly hydrophobic, so that the molecules are insoluble 
in water, Langmuir monolayers form a quasi-two dimensional system. This 
system can be described in terms of the surface pressure and temperature and  
it displays several phases with different structural properties (see, for 
instance Ref \cite{kaganer}).      
   
A chiral molecule exists in two forms $+$ and $-$, called enantiomers,  
related by a spatial transformation that involves a change of parity.    
An important feature of the physics of chiral Langmuir monolayers is the  
determination of the chiral discrimination, which occurs when     
the interaction energy between enantiomers with the same chirality is  
different from the interaction energy of enantiomers with different  
chirality. When the intermolecular forces favor the attraction between  
enantiomers of the same species, they are denominated homochiral and they  
lead to chiral segregation. On the contrary, if the attraction between  
different enantiomers is favored,  
they are named heterochiral and they lead to a racemic mixture.  
  
To study the effect of the chirality theoretically,   
Andelman proposed a simple lattice   
gas model that can be described by the Hamiltonian \cite{pelizzola}   
\begin{equation}   
{\cal H} = - \sum_{<i,j>} \sum_{r,s} \epsilon_{r  
s}N_{r,i}N_{s,j} - \sum_{i} \sum_{s}\mu_s N_{s,i} ,  
\label{e1}   
\end{equation}   
where the first sum is over nearest-neighbor pairs, the letters $i$ and $j$  
denote the sites of a two-dimensional triangular lattice, the letters $r$  
and $s$ denote the enantiomer species ($r,s = +$ or $-$), $\epsilon_{rs}$  
are the coupling energies ($\epsilon_{++} = \epsilon_{--}$ and  
$\epsilon_{+-} = \epsilon_{-+}$), $N_{r,i} = 0,1$ are the occupation  
numbers at site $i$, and $\mu_s$ is the chemical potential of the species
$s$.   
   
This model is equivalent to the Blume-Emery-Griffiths (BEG) spin-1 model  
\cite{BEGMODEL,BEG2}, as it can be seen by relating the occupation numbers 
and the spin-1 variables through the relations  
\begin{equation}   
N_{+,i} = \frac{\sigma_i^2 + \sigma_i}{2}, \qquad N_{-,i} = \frac{\sigma_i^2  
- \sigma_i}{2} ,   
\label{e2}   
\end{equation}   
where $\sigma_i = 0, \pm 1$. Thus, $\sigma_i = 1 \quad (-1)$  
represents a $+$ ($-$) enantiomer and $\sigma_i = 0$ a vacancy. In this  
way, we obtain the BEG Hamiltonian   
\begin{equation}    
{\cal H} =-\sum_{(i,j)}[J \, \sigma_{i}\sigma_{j} + \phi \,  
\sigma_{i}^{2}\sigma_{j}^{2}] -\sum_{i}[H \sigma_i - \Delta \sigma_i^2] ,   
\label{e3}   
\end{equation}  
The case $J>0$ corresponds to the homochiral case (ferromagnetic BEG).  
When $J<0$, we have the heterochiral one (antiferromagnetic BEG).   
We will concern ourselves with the homochiral case, since it has not been    
studied beyond the mean-field approach. The parameters $J, \phi$ depend on
the interaction energies $\epsilon_{++} = \epsilon_{--}$ and $\epsilon_{+-} = 
\epsilon_{-+}$ between nearest-neighbor enantiomers through the formulae    
\begin{equation}   
J=\frac{1}{2}(\epsilon_{++}-\epsilon_{+-}),   
\label{e4}  
\end{equation}  
and   
\begin{equation}   
\phi=\frac{1}{2}(\epsilon_{++}+\epsilon_{+-}) .  
\label{e5}    
\end{equation}  
The fields $H$ and $\Delta$ are related to the chemical potential of the 
species $+$ and $-$ and they are given by   
\begin{equation}    
{\it H} = \frac{\mu_{+}-\mu_{-}}{2},  
\label{e6}   
\end{equation}   
and   
\begin{equation}   
-\Delta = \frac{\mu_{+}+\mu_{-}}{2}.   
\label{e7}  
\end{equation}  
They are the conjugate parameters  of the chiral order parameter and the  
density of enantiomers defined by   
\begin{equation}   
M \equiv \left \langle \sum_{i=1}^{V}\sigma_{i} \right \rangle = \;  
\langle N_{+} \rangle - \langle N_{-} \rangle ,   
\label{e8}  
\end{equation}  
and   
\begin{equation}   
Q \equiv \left \langle \sum_{i=1}^{V} \sigma_{i}^{2} \right \rangle = \;  
\langle N_{+} \rangle + \langle N_{-} \rangle,  
\label{e9}  
\end{equation}  
where the $N_{\pm}$ are the total number of enantiomers $\pm$ and $V=L^2$ 
is the number of lattice sites. In particular, we are interested in 
determining the concentration of enantiomers $+$ or $-$ ($x_{+}$ or $x_{-}$,  
respectively) given by   
\begin{equation}  
x_{\pm}=\frac{\langle N_{\pm} \rangle}{\langle N_{+} \rangle + \langle N_{-}  
\rangle} = \frac{1}{2}(1 \pm \frac{M}{Q}) = \frac{1}{2}(1 \pm \frac{m}{q}), 
\label{e10}  
\end{equation}  
where $m = M/V$, $q = Q/V$.  
 
In this work we 
shall study first-order transitions between {\em concentrated phases} 
where the enantiomers are close to each other (phases $C_+$ and $C_-$ rich in  
enantiomers of type $+$ and $-$, respectively) and the so called {\em liquid 
expanded} ($LE$) {\em phases}, where there are many vacancies. In the spin-1 
language, we shall study transitions between ferromagnetic and 
paramagnetic phases rich in zero spins.  
 
  
\section{Monte Carlo method}   
  
\subsection{Cluster algorithm}  
  
\begin{figure}      
\includegraphics[scale=0.36]{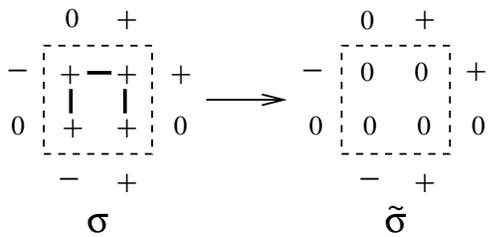}
\caption{Example of a possible cluster transition. The heavy lines are the  
activated links.}   
\label{fig1}   
\end{figure}

In experiments involving chiral Langmuir monolayers the non chiral  
contribution to the interaction energy between the enantiomers    
is usually larger than the chiral one. In our simplified model, this  
corresponds to choosing the parameter $\phi$ larger than $J$.    
In this paper, we will consider the ratio $\phi/J=3$. This choice has  
also been  previously made by Andelman \cite{andelman1} and Pelizzola et al  
\cite{pelizzola}. For $\phi/J=3$, Eq. (\ref{e3}) can be    
rewritten, up to a constant term, in the following way   
\begin{equation}   
\beta {\cal H}=-2K\sum_{<i,j>}\delta_{\sigma_{i},\sigma_{j}}+   
(\bar \Delta -2Kz)\sum_{i=1}^{N} \sigma_{i}^2-\bar H \sum_{i=1}^{N}  
\sigma_{i},   
\label{e11}    
\end{equation}  
where we used the following identity  
$-(\sigma_{i}\sigma_{j}+\sigma_{i}^{2}\sigma_{j}^2)-2(\sigma_{i}^{2}-1)    
(\sigma_{j}^{2}-1)\equiv -2\delta_{\sigma_{i},\sigma_{j}}$, $K \equiv 
\beta J$, $\bar \Delta\equiv \beta \Delta$, $\bar H \equiv \beta H$, and 
$z$ is the coordination number. For this Hamiltonian, we propose the
following cluster algorithm:    
\begin{enumerate}   
\item Choose randomly a site on the lattice and denote $\sigma_{\rm seed}$ 
 the value of its spin. This is the first spin of the cluster 
 (seed). 
  
\item Choose, with the probability 0.5, one of the two other possible spin  
 values that are different from $\sigma_{\rm seed}$. Call this new value  
 $\sigma_{\rm new}$ (it will remain fixed during the construction of the  
 cluster). For example, if $\sigma_{\rm seed}$ is $+$, $\sigma_{\rm  
 new}$ can be $-$ or $0$.  
     
\item Activate the links between the seed and its nearest neighbors that   
 are equal to $\sigma_{\rm seed}$ with probability $p=1-e^{-2K}$. Each new  
 spin connected to the cluster by an activated link is added   
 to the cluster. Next, we repeat the activation procedure to all the new
 spins   
 of the cluster. The process stops when all nearest neighbors have been  
 tested and no new spin is accepted. Now, we attempt to change this cluster  
 with spins equal to $\sigma_{\rm seed}$ into a cluster with spins  
 $\sigma_{\rm new}$ (see Fig. \ref{fig1}, for an example of a $+ 
 \rightarrow 0$ transition).  
 
\item Evaluate the difference $\delta  {\mathcal H_{\rm  bulk}}=  
 {\tilde \mathcal H_{\rm bulk}}-{\mathcal H_{\rm  bulk}}$, where   
 $\tilde \mathcal H_{\rm bulk}$ is the cluster bulk energy  
 (calculated neglecting boundary links) when all spins are  
 equal to $\sigma_{\rm new}$ and $\mathcal H_{\rm  bulk}$ is the cluster bulk  
 energy when all spins are equal to $\sigma_{\rm seed}$.    
 If $\delta \mathcal H_{\rm bulk} \le 0$, we change all spins in the  
 cluster to $\sigma_{\rm new}$ with probability $P_{\rm flip}(\sigma  
 \rightarrow \tilde \sigma) = 1$. If  
 $\delta \mathcal H_{\rm bulk} > 0$, we change all spins in the cluster to  
 $\sigma_{\rm new}$ with probability $P_{\rm flip}(\sigma \rightarrow  
 \tilde \sigma) = \exp(-\beta \delta \cal H_{\rm bulk})$.    
\end{enumerate}   
    
To prove that the algorithm satisfies the detailed balance  
condition we have to consider two types of transitions:  $\pm  
\leftrightarrow \mp$ and $ \pm \leftrightarrow 0$.     
For the first transition our algorithm is equivalent to  
Wolff's \cite{wolff} and for this reason we shall concentrate on  
transitions of the second type. Let us consider, to exemplify, the  
transition  $+ \leftrightarrow 0$. From  Eq. (\ref{e11}), we obtain    
\begin{equation}    
\frac{e^{\beta {\cal H}}}{e^{\beta \tilde{\cal H}}}=\frac   
{e^{-2K\ell_{++}}}{e^{-2K \tilde \ell_{00}}} e^{-\beta \delta {\mathcal  
H_{\rm  bulk}}},   
\label{e12}    
\end{equation}  
where $\ell_{\alpha \gamma}$ is the total number of boundary links    
that connect sites with spins $\alpha$ inside the cluster and sites with  
spins $\gamma$ outside the cluster.   
  
The ratio between the transition probability $W_{\sigma \rightarrow \tilde  
\sigma}$ and the reverse transition probability $W_{\tilde \sigma  
\rightarrow \sigma}$ is given by   
\begin{equation}    
\frac{W_{\sigma \rightarrow \tilde \sigma}}{W_{\tilde \sigma \rightarrow  
\sigma}} = \frac{ w_{\rm bulk}(1-p)^{\ell_{++}}P_{\rm flip}(\sigma  
\rightarrow \tilde \sigma)}{\tilde w_{\rm bulk}   
(1-p)^{\tilde \ell_{00}} P_{\rm flip}(\tilde \sigma \rightarrow \sigma)}.  
\label{e13}   
\end{equation}  
The bulk term $w_{\rm bulk}$ is the sum of the probabilities associated   
with all possible ways of activating links to construct the cluster, the 
term $(1-p)^{\ell_{++}}$ is the probability of not including in the cluster a  
nearest neighbor site with occupation variable $\sigma_{\rm  
seed}$. Analogous comments hold for the transition $W_{\tilde \sigma 
\rightarrow \sigma}$.     
Clearly, $w_{\rm bulk}=\tilde w_{\rm bulk}$ because for each configuration  
of activated links in $\sigma$ there is a corresponding one in $\tilde  
\sigma$ (see Fig. \ref{fig1}). Recalling the definition of $P_{\rm  
flip}$, given in step 4 of the algorithm, we see that the ratio of the  
flipping probabilities is always equal to $\exp (-\beta \delta \mathcal  
H_{\rm bulk})$. Finally, since $1-p=e^{-2K}$ the  
right hand sides of Eqs. (\ref{e12}) and (\ref{e13}) are equal  
and this equality implies detailed balance. It is worth mentioning that
the algorithm proposed here is a particular case     
of the cluster algorithm proposed by Bouabci and Carneiro \cite{bouabci}    
and later extended by Rachadi and Benyoussef \cite{rachadi} for other  
regions of the parameter space.    
  
  
\subsection{The  Sauerwein and de Oliveira method}  
  
In order to determine the grand canonical potential from Monte Carlo  
simulations, one usually  
calculates one of its derivatives and numerically integrate the  
results. To use this technique one has to know the value of the grand 
canonical potential   
at a reference point and then numerically integrate along a path which  
connects the reference point to the point where one wants to calculate the  
grand canonical potential. An alternative is the  
method proposed by Sauerwein and de Oliveira \cite{sauerwein} that allows  
one to directly obtain the grand canonical potential from the MC simulation.  
   
In this method,  the largest  
eigenvalue of the transfer matrix is  directly evaluated from Monte Carlo  
simulations. Since in the thermodynamic limit the grand partition function is  
proportional to the largest eigenvalue of the transfer matrix, its  
calculation enables us to determine all thermodynamic properties, in
particular the surface pressure that is the negative of the grand canonical 
potential. 
   
In order to explain how to obtain the largest eigenvalue, let us
consider a triangular lattice with $V$ sites divided in $N$ successive 
layers     
$S_k \equiv (\sigma_{1,k},\sigma_{2,k},...,\sigma_{L,k})$ with $L$  
spins, $V = L \times N$ (All this applies to the   
triangular lattice that we use in our paper.). The Hamiltonian may    
be decomposed in the following way   
\begin{equation}   
{\cal H}= \sum_{k=1}^N {\cal H}(S_k,S_{k+1}),   
\label{e14}  
\end{equation}  
where due to the periodic boundary conditions $S_{N+1} = S_1$.  
The probability $P(S_{1},S_{2},...,S_{N})$ of a given configuration    
of the system is given by   
\begin{equation}    
P(S_{1},S_{2},...,S_{N})=\frac{1}{Z}   
T(S_{1},S_{2}) T(S_{2},S_{3})...T(S_{N},S_{1}),   
\label{e15}  
\end{equation}  
where $T(S_{k},S_{k+1}) \equiv \exp ( - \beta {\cal H}(S_k,S_{k+1}))$ is an  
element of the transfer matrix $T$ and     
\begin{equation}   
Z=\rm Tr(\it {T^{N}}),   
\label{e16}  
\end{equation}   
is the grand-canonical partition function. By using the spectral
expansion of the matrix $T$ it is possible to show \cite{sauerwein} that  
\begin{equation}  
<\delta_{S_{1},S_{2}}> = \frac{1}{\lambda_0} <T(S_{1},S_{1})> .  
\label{e17}   
\end{equation}  
This expression enables us to calculate the largest eigenvalue $\lambda_0$
of the transfer matrix $T$ in terms of the   
averages $<\delta_{S_{1},S_{2}}>$ and $<T(S_{1},S_{1})>$, where
$\delta_{S_{1},S_{2}} = 1$ when layers $S_{1}$ and $S_{2}$ are
equal and zero otherwise. We use a MC simulation to generate the
configurations with which we calculate the averages.
 
In the specific case of the BEG Hamiltonian in the triangular lattice  
with $L \times L$ sites,  
the transfer matrix $T$ of a $n$-layer is given by  
\begin{eqnarray}  
&&T( S_{n},S_{n+1}) = \exp\{\sum_{k=1}^{L}  
[K\sigma_{k,n}(\sigma_{k,n+1}+  
\sigma_{k+1,n}\nonumber \\  
&& + \sigma_{k+1,n+1}) + \beta \phi \sigma_{k,n}^{2}(\sigma_{k,n+1}^{2}  
+ \sigma_{k+1,n}^{2}  
+ \sigma_{k+1,n+1}^{2}) \nonumber \\  
&& - \bar \Delta \sigma_{k,n}^{2}+\bar H\sigma_{k,n}]\}.  
\label{e18}  
\end{eqnarray} 
The grand canonical potential per site in the lattice gas
representation (or the free energy in the spin-1 representation) is
given by   
\begin{equation}  
\psi=-\frac{1}{\beta L} \ln \lambda_{0} = -{\cal P},  
\label{e19}  
\end{equation} 
where ${\cal P}$ is the surface pressure.  

  
\section{Numerical results}   
 
In this section, we define the following dimensionless quantities: 
\begin{equation}  
t \equiv k_B T / J , \quad D \equiv \Delta / J , \quad h \equiv H / J , 
\quad \Pi \equiv {\cal P} / J , 
\label{e20}  
\end{equation} 
where ${\mathcal P}$, the surface pressure, is given by Eq. (\ref{e19}).    
   
\begin{figure}      
\includegraphics[scale=0.36]{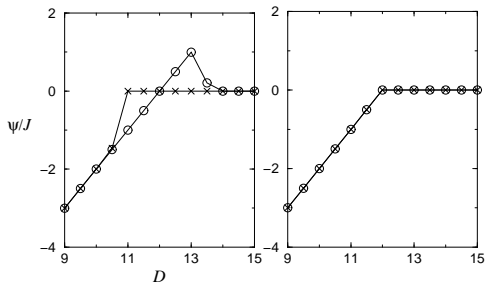}
\caption{Grand canonical  potential $\psi/J$ versus $D$ across   
the first-order line for $L=30$, $h=0$ and $t=0.8$.    
The first graph refers to the Metropolis algorithm and the second   
to the cluster algorithm. The symbol $\circ$ ($\times$)  
indicates increasing (decreasing) $D$. The symbols are larger than the
error bars.} 
\label{fig2}   
\end{figure}

As a check on the efficiency of the proposed cluster algorithm, we    
show in Fig. \ref{fig2} the grand canonical potential $\psi$   
versus the chemical potential $D$ for $h=0$ and $t=0.8$. We  
considered a very low temperature, because   
in this case hysteresis effects are very strong. In  
Fig. \ref{fig2} we compare the performances of the Metropolis and the  
cluster algorithm on a triangular lattice with periodic boundary conditions
and linear dimension $L=30$. To evaluate $\psi$ and to
estimate its statistical error after
equilibrating the systems we have used $5 \times 10^{4}$ Monte  
Carlo steps divided into $1000$ independent runs. Note
that with the Metropolis algorithm the system is trapped    
in metastable states and even after millions of MC steps it does not undergo   
a transition to the stable phase.    
This does not happen with the cluster algorithm because  
the system is able to easily pass from one phase to the other. The
efficiency of the algorithm allows us to     
determine first-order transition lines with high precision and  
the good quality of the data enables us to perform very precise   
finite size analysis.

In principle it is possible to determine the transition point
using the free energy. As it can be seen in Fig. \ref{fig2}, there is a
kink in the free energy as a function of $D$ at the transition point
$D_{L}^{*}$. One can then perform a finite size analysis to obtain
$D_{\infty}^{*}$. However, it is simpler and more efficient to analyze the
susceptibility whose
finite size behavior is well known for both first and
second-order phase transitions. After determining $D_{\infty}^{*}$ we use the
SO method at this point to calculate the surface pressure. In first-order
phase transitions, the surface pressure, which is proportional to the
negative of the grand canonical potential, does not have a finite size
behavior as simple as the susceptibility. As a 
function of the system size, the surface pressure saturates quickly. Thus,
the values of the surface pressure that we use in our graphs come from the 
largest lattices that we have simulated.     
    
\begin{figure}  
\includegraphics[scale=0.36]{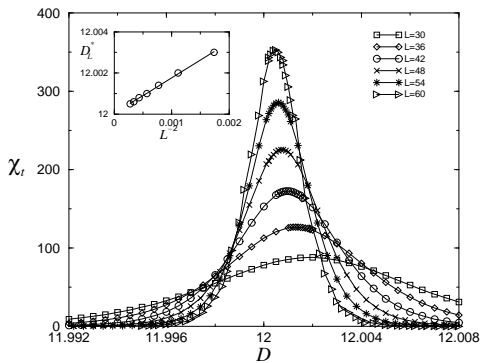}   
\caption{Susceptibility $\chi_{t}$ versus $D$ for several  
values of system size $L$, $h=0$ and $t=2.4$. In the inset, we plotted the 
value of $D$ for which the susceptibility is maximum ($D_{L}^{*}$) versus 
$L^{-2}$.}   
\label{fig3}   
\end{figure}

The susceptibility is defined  
as $\chi_{t}= L^{2}(\langle m^{2} \rangle-\langle |m|   
\rangle^{2}) / t$, where the magnetization $m = \sum_i \sigma_i  / V$.    
For a fixed system size $L$, maintaining $t$ and $h$ fixed, and   
increasing $D$ towards the coexistence line,    
one observes a peak in the susceptibility at $D_{L}^{*}$, as seen in
Fig. \ref{fig3}, where the lines were drawn only to guide the eye. In   
the thermodynamic limit, this peak becomes a delta-function singularity.    
According to Refs. \cite{rBoKo,challa}, the deviation of $D_{L}^{*}$ from  
its asymptotic value $D_{\infty}^{*}$ decays as $L^{-2}$, in agreement with  
our results, shown in the inset of Fig. \ref{fig3} . From this law, we
have obtained the extrapolated value $D_{\infty}^{*}=12.0000(1)$.
  
\begin{figure}      
\includegraphics[scale=0.36]{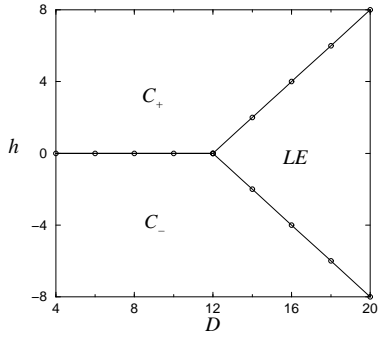}
\caption{Phase diagram in the space of the chemical potentials   
$h$ versus $D$. The symbols $C_+$ and $C_-$ denote the condensed phases 
rich in enantiomers $+$ and $-$, respectively, and $LE$ is the liquid 
expanded phase. The solid line is the $t=0$ calculation which practically 
coincides with the mean field result. The circles are from MC
simulations.}    
\label{fig4}   
\end{figure}

In order to understand this result let us perform an exact zero
temperature calculation of the transition point $D_{L}^{*}$. At zero
temperature the free energy $F=U - TS = U \equiv <{\mathcal H}>$, where
${\mathcal H}$ is given in Eq. (\ref{e3}). The system chooses the phase
that minimizes the energy $U$. At $T=0$, for small values of $D$ all spins are
$+1$ if $h > 0$ or $-1$ if $h < 0$. If $D$ is large enough, the energy is
minimized when all spins are $0$. The transition line is obtained by
equating the energies in the ferromagnetic (condensed) phase with all
$\sigma_i = +$ (or $-$) and in the paramagnetic (liquid expanded) phase
with all $\sigma_i =0$. Taking into 
account Eq. (\ref{e3}), with $\phi = 3J$, and the definitions given in
Eq. (\ref{e20}), we calculate the energy $U_\pm$ of the ferromagnetic phases  
and the energy $U_0$ of the paramagnetic phase 
\begin{equation}
U_\pm = VJ(-12  \mp h + D)\,; \quad U_0 = 0.   
\label{e21}  
\end{equation} 
The equation $U_\pm = U_0 \Leftrightarrow D = 12 \pm h$ gives the
transition lines between the ferromagnetic and paramagnetic phases.
The equation $U_+ = U_- \Leftrightarrow h = 0$ gives the transition line
between the two ferromagnetic phases (this holds for $D \leq 12$, for
$D>12$ the systems is in the paramagnetic 
phase). All these transition lines are
represented in Fig. \ref{fig4} that gives the phase diagram of the Langmuir
monolayer in the plane of the chemical potentials $h \times D$ (Recall
that $hJ = H = (\mu_+ - \mu_-)/2$ and  $DJ = \Delta = (- \mu_+ -
\mu_-)/2$.). The circles are the results of Monte Carlo simulations
performed at $t=2.4$. The error bars are 
smaller than the circles. It is 
interesting to remark that in the temperature interval relevant for Langmuir
monolayers the
zero temperature calculations give practically the same results as the MC
simulations and the mean-field calculations for the transition lines. 
In Langmuir monolayers language, for
$h=0$ and low values of $D$ (higher chemical potentials),    
we have the condensed phase characterized    
by a $1:1$ mixture of the two enantiomers. For higher values of $D$ a    
transition from the condensed phase, rich in enantiomers $\pm 1$, to   
the phase poor in enantiomers, the  liquid expanded phase, takes place.  
When the chemical potential of the species are different ($h \neq 0$), we 
have larger fraction of   
enantiomers $+$ $(-)$ whenever $h>0$ $(h<0)$, and in the  limit of   
$h>>0$ $(<<0)$ the solution only contains the enantiomer $+$ $(-)$.   

\begin{figure}      
\includegraphics[scale=0.36]{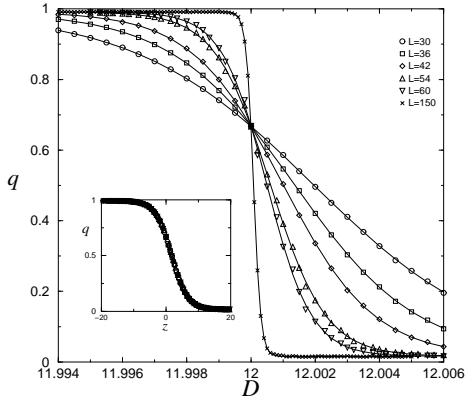}
\caption{Order parameter $q$ versus $D$ for $h=0$, $t=2.4$ and several system  
sizes $L$. In the inset, a collapse of all curves by plotting $q$ versus 
$z=(D-D_{\infty}^{*})*L^{2}$.}    
\label{fig5}   
\end{figure}

Another procedure for locating the phase transition   
consists in determining the crossing point of the $q$ versus $D$ isotherms 
for different system sizes.   
As showed in the Ref. \cite{fiore4}, the crossing point is independent   
of the lattice size and properly identifies  
the transition, as shown in  Fig. \ref{fig5}. We shall present below an
independent derivation of this important result based on the work of Borgs and
Koteck\'y \cite{rBoKo}. 
      
For $h=0$ all curves of $q$ versus $D$ cross at $D^{*} =
12.0000(1)$ and $q \approx 2/3$ for this value of $D$. This criterion for
estimating the value of $D$ for which the phase transition takes place
agrees very well with the finite size analysis of the susceptibility
$\chi_{t}$ that we have discussed above. For $h \neq 0$,  
two phases coexist at the point $D_{h}^{*}$ which now depends on $h$ and  
all isotherms cross at  $q \approx 0.5$. We remark that if single flip  
algorithms are used to generate the dynamics, one will not be able to 
determine the crossing of the curves due to hysteresis effects.  
  
More relevant from the point of view of Langmuir monolayers,   
and other physical systems involving mixture of molecules,  
are the surface-pressure versus concentration diagrams. But before 
discussing this phase diagram  
we will describe our procedure to fit the curves in Fig. \ref{fig5} and to
obtain the $V \rightarrow \infty$ limit  
of $q$ and $m$ that are used to determine the concentration $x_+$
(see Eq. (\ref{e10})).  
Since the simulated system is finite, the calculated  
quantities will be affected by finite size effects. As mentioned  
previously, in the last  
years, the finite size theory of first order phase transitions has been  
studied extensively for quantities, such as the specific heat and the  
susceptibility. There are fewer studies for the dependence on the  
system size of quantities like the magnetization or the concentration of  
molecules \cite{gupta,borgs94}. In the following, we propose a method   
to determine the concentrations of the phases that  
coexist directly from the numerical  
simulations.  The first step consists in noting  
that $q \times D$ (or $m \times D$) isotherms can be fitted by the  
equation  
\begin{equation}  
q = \frac{b+c e^{-a \,\delta D}}{1+d e^{-a \,\delta D}},  
\label{e22}  
\end{equation}  
where $a$, $b$, $c$ and $d$ are fitting parameters and $\delta D \equiv
D-D^{*}_{\infty}$. We are going to show below that $a$ depends on the system
size $L$ and the temperature $T$.   
An analogous expression can be written down for the order parameter $m$.  
The expression above was inspired by the work of Borgs and Koteck\'y  
\cite{rBoKo}, where it is shown that at low temperatures the partition  
function for two coexisting phases can be written as   
\begin{equation}  
Z = [ e^{-\beta f_1(\beta,h) V} + e^{-\beta f_2(\beta,h) V} ](1 +  
e^{-L/L_0}) ,   
\label{e23}  
\end{equation}  
where $h$ is the magnetic field (our system also depends on the crystal  
field $D$), $L_0$ is a constant of the order of the infinite volume 
correlation length and $f_i$ is a metastable free energy 
for the phase $i$ ($i = 1 \; \mbox{or} \; 2$). 

In our system, for $h=0$ three phases coexist at the triple point
  ($D_\infty^* = 12$). Thus,   
we expect the sum of three exponentials instead of two as in Eq. (\ref{e23}).  
We have assumed that all three exponentials have the same weight (we shall
use our results to check this point). In the neighborhood of the
triple point 
\begin{equation}  
Z \approx e^{-\beta V f_0} + e^{-\beta V f_+} + e^{-\beta V f_-} ,
\label{e24}  
\end{equation}  
where the $f_i=f_i(\beta,h,D)$, $i = 0,\pm$, are respectively the
metastable free energies of the paramagnetic and ferromagnet phases. Away
from the coexistence curve, only the $f_i$ associated with the correct
phase remains and becomes the 
free energy of the system ($Z = \exp (-\beta V f_i) $ ).

The parameters $m$ and $q$ are given by 
\begin{equation}  
q = - \frac{1}{\beta V} \frac{\partial \log Z}{\partial D}, \quad m = -
\frac{1}{\beta V} \frac{\partial \log Z}{\partial h} .
\label{e25}  
\end{equation}  

For $h=0$, $f_+ = f_- \equiv f_\pm$. Taking into account this fact and
Eqs. (\ref{e24}) and (\ref{e25}), we can write the parameter $q$ as 
\begin{equation}  
q \approx \frac{(\partial f_0 / \partial D)e^{-\beta V f_0} +
  2 (\partial f_\pm / \partial D) e^{-\beta V f_\pm}}{e^{-\beta V f_0} +
  2 e^{-\beta V f_\pm}} . 
\label{e26}  
\end{equation}

At the triple point $f_0^* = f_\pm^*$, where $f_i^* \equiv
f_i(\beta,h=0,D=D_\infty^*)$, $i = 0,\pm$  and the exponentials in
Eq. (\ref{e26}), which contain the only dependence on the lattice size,
cancel out and we obtain 
\begin{equation}  
q^* \equiv q(\beta,0,D_\infty^*) \approx \frac{1}{3} \left [ 
\left . \frac{\partial f_0}{\partial D} \right |_{D=D_\infty^*} + 
\left . 2 \frac{\partial f_\pm}{\partial D} \right |_{D=D_\infty^*}  \right ].
\label{e27}  
\end{equation}  
This is the reason why the  $q \times D$ curves
for different lattice sizes cross at the same point. The crossing point
provides another method do locate the phase boundaries. 

Our calculations are performed at low temperatures. In mean-field, the
temperatures are measured in units of the coordination number $z$. For the
triangular lattice, $z=6$. Our MC temperature $t=2.4$ is equivalent
to a $t=0.4$ mean-field temperature. We can use the exact zero
temperature energies given in Eq. (\ref{e21}) to estimate the derivatives
in Eq. (\ref{e27}). Recalling that $f_{i} = U_{i} / V J$, $i =
0,\pm$, at $t=0$, we 
obtain $\partial f_0 / \partial D = 0$ and $\partial f_\pm / \partial D = 
1$. Thus, at the triple point $q^* \approx 
2/3$ which is the result that we obtain in Fig. \ref{fig5}.

An analogous demonstration holds for the case $h \neq 0$, where $Z$ is the
sum of two exponentials as in Eq. (\ref{e23}). The factor $2$ in
Eq. (\ref{e29}) is replaced by $1$ and the
crossing of the $q \times D$ curves occurs at the point $D \approx 0.5$.

In the curves plotted in Fig. \ref{fig5}, $D$ varies in the interval
$[11.994,12.006]$ which is very narrow. It is possible in this case to
expand the $f_i=f_i(\beta,h=0,D)$, $i = 0,\pm$, around the triple point, 
\begin{equation}  
f_i = f_i^* + f_i^{\,\prime *}\, \delta D + {\mathcal O}((\delta D)^2),
\label{e28}  
\end{equation}
where $\delta D \equiv D-D_\infty^*$, $f_i^* \equiv
f_i(\beta,h=0,D=D_\infty^*)$ and $f_i^{\,\prime *} \equiv (\partial f_i /
\partial D)|\{D=D_\infty^*\}$, for $i = 0,\pm$, 
\begin{equation}
q \approx
\frac{f_0^{\,\prime *}\,e^{-\beta V f_0^{\,\prime *}\,\delta D} +
  2 f_\pm^{\,\prime *}\,e^{-\beta V f_\pm^{\,\prime
  *}\,\delta D}}{e^{-\beta V f_0^{\,\prime *}\,\delta D} + 
  2 e^{-\beta V f_\pm^{\,\prime *}\,\delta D}} ,   
\label{e29}  
\end{equation}  
which has the same form as Eq. (\ref{e22}) after we divide the numerator
and the denominator by $\exp (-\beta V f_0^{\,\prime *}\,\delta D)$. 

In Fig. \ref{fig5} the symbols stand for the  
values of $q$ obtained from the numerical simulations    
and the solid lines are fits of the points using Eq. (\ref{e22}) by
minimizing the $\chi^2$ merit function \cite{press}.  In  
order to perform the fittings we used the Levenberg-Marquardt method that
is well described in Ref. \cite{press}, where one can also find the
subroutines that are necessary to implement the method. These subroutines
return the variances of the fitting parameters and the  
quality of the fitting. A few words about the implementation of the
subroutines is in order. Our fitting function Eq. (\ref{e22}) contains
exponentials whose arguments may become very large. In order to avoid
numerical overflow it is convenient to use the asymptotic values of $q$
when $|\delta D|$ becomes too large. Define, for example,
$q \equiv b$ for $\delta D > 30$ and  $q \equiv c/d$ for $\delta D< -30$. Of
course, the number $30$ is rather 
arbitrary. Non-linear fittings depend on a good initial guess of the
fitting parameters. One may proceed as follows. Note that 
$b = q(D \rightarrow \infty)$ and $c/d = q(D \rightarrow - \infty)$. Since
in the simulations the $D$ interval is finite, instead of taking the $|D|
\rightarrow \infty$ limit we use the values of $q$ in our data set
associated with the largest and the smallest values of $D$. Call them
$q_+$ and $q_-$, respectively and put $b \approx q_+$, $c/d \approx
q_-$. Next define $q^* \equiv q(D=D_\infty^*)$ and $q_1 \equiv q(D =
D_1)$, where $D_1 <  D_\infty^*$, is chosen in the region where the graph
$q \times D$ has already started to curve down. It is simple
to solve the fitting parameters in terms of these quantities.
\begin{eqnarray}
&& a = \frac{1}{D_\infty^* - D_1}\log \left | \frac{(q_1 - q_+)(q_- -
  q^*)}{(q_- - q_1)(q^* - q_+) }\right |, \nonumber \\
&& b = q_+, \quad c = \frac{q_-(q^* - q_+)}{q_- - q^*}, \quad d =
  \frac{q^* - q_+}{q_- - q^*}. 
\label{e30}
\end{eqnarray}
With this choice for the initial parameters, the convergence of the fitting
routine is very fast and the quality of the 
fitting is very good (the factor $Q$ that measures the goodness-of-fit
\cite{press} is close to $1$). In Tables \ref{table1} and \ref{table2} the
errors of the parameters are the square roots of the variances (standard
deviations) that are returned by the fitting routines.  

The fitting parameters for the curves in  Fig. \ref{fig5} are given in
Table \ref{table1}. Now we can check the equal weight hypothesis for the
exponentials. For $h=0$ the two condensed phases $C_\pm$ have the same free  
energy and two of the three exponentials are identical, as we discussed
above. The $q \times D$ curves in Fig. \ref{fig5} are in the
vicinity of the triple point, so    
we expect that $d \approx 2$. This is the result that we 
obtain (see Tables \ref{table1} and \ref{table2} for the magnetization $m$). 

For $h \neq 0$, there 
is the coexistence of two phases ($LE$ and $C_+$ or $C_-$). Near
the transition we have the sum of two exponentials with the same weight,
as in Eq. (\ref{e23}). We have checked that $d \approx 1$ near the transition
line, as it was expected. 
   
Comparing Eqs. (\ref{e22}) and (\ref{e26}) we note that the parameter $a$ that
appears in the exponent should be proportional to the system volume.    
A $\log(a) \times \log(L)$ plot gives the straight  
line $\log(a) = A + B \log(L)$ with $A = -0.87(2)$ and $B = 1.990(6)$ for
table \ref{table1}; and $A = -0.88(2)$ and $B = 1.993(5)$ for  
table \ref{table2}. Thus, as expected, the constant $a$ scales with the  
volume of the system.  
  
\begin{figure}       
\includegraphics[scale=0.36]{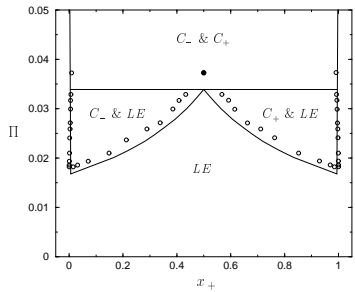}
\caption{Surface pressure $\Pi$ versus concentration $x_{+}$ phase diagram  
obtained by numerical simulations (circles) and mean field (lines).}   
\label{fig6}   
\end{figure}

The values of $q$ for the condensed and liquid expanded phases  
are calculated by taking the $L \rightarrow \infty$ limit in  
Eq. (\ref{e22}). The condensed phases occur in the region $D - D_\infty^*
< 0$ and for this reason $q \rightarrow c/d$ as $L \rightarrow \infty$. The  
liquid expanded phase occurs in the region $D - D_\infty^* > 0$ and $q   
\rightarrow b$ as $L \rightarrow \infty$. The curves for $m \times D$ are
very similar to the curves $q \times D$ in Fig. \ref{fig5} and can be also be  
fitted by an expression analogous to Eq. \ref{e22}. Having calculated $q$  
and $m$, we obtain $x_{+}$ through the expression $x_{+} = (1
+|m|/q)/2$. The use of $|m|$ instead of $m$ is due to technical reasons
(see sections 2.3.3 and 2.3.4 in Ref. \cite{binder}). As a consequence,
the magnetization is small but not zero when $h=0$. This introduces a
small distortion in the diagram of Fig. \ref{fig6} near $x_+ = 0.5$, but
symmetry arguments guarantee that $m=0$ when $h=0$ and the coexistence curve 
passes through the point with $x_+ = 0.5$ (filled circle in
Fig. \ref{fig6}). 

The $h>0$ half side of the diagram $h \times D$ is mapped onto the right  
hand side of the   
surface-pressure versus concentration diagram ($x_{+}>0.5$) whereas  
$h<0$ corresponds to the $x_{+}<0.5$ concentration range.   
As mentioned above,  $h=0$ implies that   
the fraction of enantiomers $+$ and $-$  
are equal and in the coexistence of the three phases, one has $x_{+}=0.5$.  
From the point of view of homochiral Langmuir monolayers,   
the chiral segregation  
takes place, in contrast to the heterochiral case, in which  
one has a racemic mixture.   
The surface pressure of a 1:1 mixture of enantiomers   
is higher than the pressure for pure enantiomers. This feature is verified  
in experiments in which the chiral segregation occurs \cite{kaganer}.  
Unfortunately, to date few experiments have been performed covering the whole  
range of concentrations, usually they are restricted to the 1:1 mixture 
and the pure cases. For comparison, we have also plotted in
Fig. \ref{fig6} the results   
obtained from the mean field technique. 
 
In contrast to the heterochiral case, for which the mean field results
disagree with those obtained from the cluster variational method, in the
homochiral case the accordance between mean field and the numerical
simulations is very good.    
  
  
\section{Conclusions}   
  
In this paper, we present an efficient way for determining   
phase diagrams from numerical simulations. To illustrate it,  
we have considered a simple model  
that describes the behavior of homochiral Langmuir  
monolayers, which is equivalent to the BEG model. It is worth  
mentioning that although we  
have interpreted the phase diagrams obtained here in terms  
of Langmuir monolayers, similar phase diagrams are obtained  
when one uses the BEG model   
to describe a mixture of two distinct species with vacancies.    
The use of a cluster algorithm  
that eliminates metastability in first order phase transitions  
allows us to precisely locate the first-order transitions lines.  
To determine the surface pressure we  
used the method proposed by Sauerwein and de Oliveira in which  
the surface pressure is determined directly from the numerical  
simulations without the necessity of performing numerical  
integrations. The fitting procedure proposed in this paper to determine
the concentrations, based on the work of Borgs and Koteck\'y \cite{rBoKo},
is easy to implement and uses all information contained 
in the order parameter curve. It seems to improve on the usual finite size
analysis for the magnetization near first-order transition lines in that
it does not present ``overshooting'' effects \cite{gupta,borgs94} and
both $m$ and $q$ present a monotonic behavior as a function of $L$, but
this point has to be further investigated by increasing the
statistics. The elimination of metastability also enables us to use the
crossing of the curves $q \times D$ (or $m \times D$) for different
lattice sizes as a criterium for locating the phase boundaries. This
usually cannot be done due to hysteresis effects. Finally, we remark that
our approach      
is general and it can be used for any spin model. In systems for which  
a cluster algorithm is not available, we can use other techniques, such as
the multicanonical approach \cite{berg} or the simulated tempering
\cite{parisi} to generate the dynamics.   
  
   
\section*{ACKNOWLEDGMENT}   
C. E. F. acknowledges the financial support from    
Funda\c c\~ao de Amparo \`a Pesquisa do   
Estado de S\~ao Paulo (FAPESP) under Grant No. 06/51286-8.  
  

 
%
\begin{table}  
\begin{ruledtabular}  
\begin{tabular}{ccccc}  
L & a & b & c & d \\  
\hline  
24 & 233.2(9) & 0.0173(7) & 1.983(8) & 2.000(8) \\  
30 & 363(1) & 0.0166(4) & 1.986(2) & 2.004(9) \\  
36 & 521(2) & 0.0161(4) & 1.99(1) & 2.01(1) \\  
42 & 708(3) & 0.0160(3) & 1.99(1) & 2.00(1) \\  
48 & 925(4) & 0.0160(3) & 1.99(1) & 2.00(1) \\  
54 & 1168(6) & 0.0160(3) & 1.98(1) & 2.00(1) \\
60 & 1.45(1)$\times 10^3$ & 0.0165(7) & 1.98(1) & 2.00(2) 
\end{tabular}  
\end{ruledtabular}  
\caption{\label{table1} Values of the fitting parameters obtained from  
the $q \times D$ curves in Fig. \ref{fig4} ($t=2.4$ and $h=0$). The 
numbers between brackets are the uncertainties in the last digits.}  
\end{table}   
\begin{table}  
\begin{ruledtabular}  
\begin{tabular}{ccccc}  
L & a & b & c & d \\  
\hline  
24 & 233.4(9) & 0.0047(6) & 1.943(8) & 1.998(8) \\  
30 & 364(1) & 0.0036(2) & 1.943(9) & 1.998(9)  \\  
36 & 523(2) & 0.0029(1) & 1.94(1) & 2.00(1) \\  
42 & 711(3) & 0.0025(1) & 1.94(1) & 2.00(1) \\  
48 & 930(4) & 0.0022(1) & 1.94(1) & 1.99(1) \\  
54 & 1175(6) & 0.00200(9) & 1.93(1) & 1.99(1) \\
60 & 1.45(1)$\times 10^3$ & 0.0018(3) & 1.93(2) & 1.99(2)
\end{tabular}  
\end{ruledtabular}  
\caption{\label{table2} Values of the fitting parameters obtained from  
the $m \times D$ curves for $t=2.4$ and $h=0$.}  
\end{table}   

\end{document}